\UseRawInputEncoding

\documentclass[10pt, prb, aps, twocolumn, showpacs, citeautoscript, floatfix, reprint, amsmath, amssymb, notitlepage,  superscriptaddress]{revtex4-1}

\usepackage{hyperref}
\usepackage{graphicx}
\usepackage{wasysym}
\usepackage{stmaryrd}
\usepackage{rotating}
\usepackage{amsmath}
\usepackage{amsfonts}
\usepackage{amssymb}
\usepackage[countmax]{subfloat}
\usepackage{dcolumn} % align table columns on decimal point
\usepackage{bm} % bold math
\usepackage{color}

\usepackage{float}
\usepackage{latexsym}

\usepackage{soul} % for strikeout fonts

\usepackage{macros}
\usepackage{braket}

\begin{document}

\title{Quantum geometry of singlet superconductors}

%\title{Revealing quantum geometric properties of singlet superconductors by infrared absorption}

%\title{Measurement of quantum metric in singlet superconductors by infrared absorption}

\author{David Porlles}
\affiliation{Department of Physics, PUC-Rio, 22451-900 Rio de Janeiro, Brazil}

\author{Wei Chen}
\affiliation{Department of Physics, PUC-Rio, 22451-900 Rio de Janeiro, Brazil}

\date{\today}

\begin{abstract}

We elaborate that $s$-wave and $d$-wave superconductors described by mean field theories possess a nontrivial quantum geometry. From the overlap of two quasihole states at slightly different momenta, one can define a quantum metric that measures the distance in the curved momentum space. The momentum-integration of the quantum metric represents an average distance that we call the fidelity number, which may be further expressed as a fidelity marker defined locally on every lattice site. For $s$-wave superconductors, we unveil that the quantum metric generally influences the electromagnetic responses at finite wave length, such as the infrared absorption and paramagnetic current. In addition, the dielectric response is directly proportional to the fidelity number, which is found to be determined by the coherence length and suppressed by disorder. For $d$-wave superconductors, we demonstrate the singular behavior of the quantum metric near the nodal points, and a metric-curvature correspondence between the azimuthal quantum metric and the non-Abelian Berry connection that integrates to a topological charge of the nodal points.

%and how it renders a diverging fidelity number. 

\end{abstract}

\maketitle

\section{Introduction}

The quantum geometry of the valence band Bloch state emerges recently as a key aspect related to various material properties of insulators and semiconductors, especially to their topological properties\cite{Ma13,Ma14,Yang15,Piechon16,
Palumbo18_2,Chen20_Palumbo,Ma20,Mera21,vonGersdorff21_metric_curvature,Chen22_dressed_Berry_metric}. Starting from the fully antisymmetric valence band Bloch state $|\psi({\bf k})\rangle$ at momentum ${\bf k}$, the notion of quantum geometry arises from considering the overlap $|\langle\psi({\bf k})|\psi({\bf k+\delta k})\rangle|=1-g_{\mu\nu}\delta k^{\mu}\delta k^{\nu}/2$ expanded in terms of the small displacement $\delta{\bf k}$, yielding a prefactor $g_{\mu\nu}$ that is referred to as the quantum metric\cite{Provost80}. The periodic Brillouin zone (BZ) is then considered as a compact Euclidean manifold equipped with this quantum metric, from which the usual quantities in differential geometry, such as Ricci scalar, Riemann tensor, geodesic, etc, can be introduced.

Besides these purely mathematical aspects, quantum metric has also been linked to various experimental measurables\cite{Ozawa18,Chen22_Cappellaro,Yu19,Tan19,Gianfrate20,Mitscherling20,Mitscherling22,Mera22}. Particularly in semiconductors, the exciton absorption rate at momentum ${\bf k}$ as a function of the frequency of a polarized light, which can be measured by detecting the loss of valence band electron population in the pump-probe type of experiments\cite{Gierz13}, is described by a quantum metric spectral function that frequency-integrates to the quantum metric\cite{Chen22_dressed_Berry_metric}. In addition, the frequency-dependence of the optical absorption rate, which has been measured in semiconductor for decades\cite{Stillman84}, as well as recently measured in 2D materials from their transmittance\cite{Nair08,Weber10,Bruna09,Nair10}, actually corresponds to the momentum integration of the quantum metric spectral function that has been called the fidelity number spectral function\cite{deSousa23_fidelity_marker,deSousa23_graphene_opacity}. The significance of this spectral function is that it frequency-integrates to a fidelity number that represents the average distance between neighboring Bloch states in the momentum space, thereby serving as a characteristic quantum geometrical property of the BZ manifold. Moreover, the fidelity number can be converted into a fidelity marker defined locally on lattice sites, pointing to the possibility of investigating the influence of real space inhomogeneity on the quantum geometrical properties of solids\cite{deSousa23_fidelity_marker}.

Besides these experimental measurable, another important feature of the quantum metric is its relation with the topological order. It has been pointed out that in systems where the topological order is given by the momentum-integration of Berry connection or Berry curvature, the module of these quantities is equal to the determinant of the filled band quantum metric\cite{Ma13,Kolodrubetz13,Ma14,Yang15,Piechon16,Kolodrubetz17,Ozawa18,Palumbo18,
Palumbo18_2,Lapa19,Yu19,Chen20_Palumbo,Ma20,Salerno20,Lin21}. Along this line of development, it is recognized recently that, in fact, the module of the curvature functions that momentum-integrate to the topological order of Dirac models in any dimension and symmetry class\cite{Schnyder08,Ryu10,Kitaev09,Chiu16} is always equal to the determinant of the quantum metric, a ubiquitous relation that has been called the metric-curvature correspondence\cite{vonGersdorff21_metric_curvature}. As a result, the aforementioned exciton absorption experiment that measures the quantum metric can help to reveal the topological order in these materials.

In addition to these remarkable features in insulating materials, quantum metric also manifests in yet another system that is currently under intensive investigation, namely the flat band superconductors (SCs). This subject rises to prominence owing to the flat band superconductivity recently discovered in twisted bilayer graphene\cite{Cao18,Yankowitz19}. Although the microscopic mechanism for the superconductivity in this system is still under intensive debate, various theories have suggested that the superfluid density therein is directly related to the quantum metric of the flat band\cite{Peotta15,Julku16,Liang17,HerzogArbeitman22,Torma22,Iskin23}.

Motivated by these intensive investigations of flat band SCs and the knowledge about optical absorption in semiconductors, in this paper we present a detailed survey on the quantum geometrical properties of the quasihole band of typical singlet SCs, including both the cases of $s$-wave and $d$-wave pairing. Our objective is to elaborate that typical singlet SCs described by Bardeen-Cooper-Schrieffer (BCS) mean field theories\cite{Bardeen57} also has nontrivial quantum geometrical properties. For $s$-wave SCs, we will elaborate the remarkably simple form of the quantum metric, and demonstrate that the metric generally appears in optical and dielectric responses. However, unlike the optical absorption in semiconductors, the infrared absorption and the so-called paramagnetic current of clean SCs is not directly given by the quantum metric spectral function owing to the complication coming from the Bogoliubov transformation, commonly known as the coherence factor\cite{Tinkham04}. On the other hand, the zero-frequency dielectric function turns out to be directly proportional to the fidelity number, which is essentially given by the coherence length measured in units of lattice constant. For $d$-wave SCs, we will emphasize the very singular momentum profile of the quantum metric, as well as the metric-curvature correspondence between the non-Abelian Berry curvature that integrates to a topological charge and the azimuthal quantum metric.

The structure of the paper is organized in the following manner. In Sec.~II, we elaborate how the notions of quantum metric, fidelity number, and fidelity marker arise from the quasihole state of singlet SCs. To further demonstrate how these quantum geometrical quantities manifest in electromagnetic responses of SCs, we proceed to lay out the general formalism for the infrared absorption, paramagnetic current, and linear screening. In Sec.~III, we turn to 3D and 2D $s$-wave SCs to illustrate their quantum geometrical properties and how they manifest in the electromagnetic responses. In Sec.~IV, the singular behavior of quantum metric in $d$-wave SC is revealed, with a special emphasis on its relation with the topological charge at the nodal points. Section V summarizes our results and discusses possible extensions of our work.

%In particular, we will use continuous models to give analytical forms of these quantities to explicitly elaborate their relation with the fidelity number.

%{\cblue (1) Actually we should make an even more profound statement for s-wave SC: Whatever quantity that is linear in coherence length is proportional to fidelity number. That sounds quite universal. }

%{\cblue (1) I should do it like Tinkham, for all the response functions, use $q\xi$ as the dimensionless parameter to expand, where the leading order term seems to be $(q\xi)^2$.  }

\section{Quantum geometry and electromagnetic responses of singlet superconductors}

\subsection{Quantum metric and fidelity marker in singlet superconductors}

We start by considering mean-field spin-singlet SCs in any spatial dimension $D$, whose single-particle Hamiltonian takes the form of a $2\times 2$ Dirac Hamiltonian 
\begin{eqnarray}
H({\bf k})={\bf d}\cdot{\boldsymbol\sigma}=d_{1}\sigma_{1}+d_{3}\sigma_{3},
\end{eqnarray}
where $\sigma_{i}$ are the Pauli matrices, $d_{1}=\Delta_{\bf k}$ is the momentum-dependent superconducting gap, and $d_{3}=\varepsilon_{\bf k}$ is the normal state dispersion. The basis of the Hamiltonian is $|\psi_{\bf k}\rangle=(c_{\bf k\uparrow},c_{\bf -k\downarrow}^{\dag})^{T}$. The ${\bf d}$-vector divided by its module defines a unit vector
\begin{eqnarray}
{\bf n}\equiv{\bf d}/|{\bf d}|=(d_{1}/d,d_{3}/d)=(n_{1},n_{3}),
\label{nvector_definition}
\end{eqnarray}
with $\pm d=\sqrt{d_{1}^{2}+d_{3}^{2}}=\pm\sqrt{\varepsilon_{\bf k}^{2}+\Delta_{\bf k}^{2}}=\pm E_{\bf k}$ the dispersion of the two bands. We denote the filled quasihole eigenstate with eigenenergy $-E_{\bf k}$ by $|n({\bf k})\rangle\equiv|n\rangle$ (not to be confused with the ${\bf n}$-vector in Eq.~(\ref{nvector_definition})), and the empty quasiparticle eigenstate with eigenenergy $+E_{\bf k}$ by $|m({\bf k})\rangle\equiv|m\rangle$, which take the form
\begin{eqnarray}
&&|n\rangle=\frac{1}{\sqrt{2d(d-d_{3})}}\left(\begin{array}{c}
d-d_{3} \\
-d_{1}
\end{array}\right)={\rm Sgn}(\Delta_{\bf k})\left(\begin{array}{c}
v_{\bf k} \\
-u_{\bf k}
\end{array}\right),
\nonumber \\
&&|m\rangle=\frac{1}{\sqrt{2d(d+d_{3})}}\left(\begin{array}{c}
d+d_{3} \\
d_{1}
\end{array}\right)=\left(\begin{array}{c}
u_{\bf k} \\
v_{\bf k}
\end{array}\right),
\label{quasihole_quasiparticle_states}
\end{eqnarray} 
where $u_{\bf k}$ and $v_{\bf k}$ are the usual Bogoliubov coefficients
\begin{eqnarray}
&&c_{\bf k\uparrow}=u_{\bf k}\gamma_{\bf k\uparrow}+v_{\bf k}\gamma_{\bf -k\downarrow}^{\dag},\;\;\;
c_{\bf -k\downarrow}=u_{\bf k}\gamma_{\bf -k\downarrow}-v_{\bf k}\gamma_{\bf k\uparrow}^{\dag},
\nonumber \\
&&u_{\bf k}=\sqrt{\frac{1}{2}\left(1+\frac{d_{3}}{d}\right)},\;\;\;v_{\bf k}={\rm Sgn}(\Delta_{\bf k})\sqrt{\frac{1}{2}\left(1-\frac{d_{3}}{d}\right)},
\nonumber \\
\label{uk_vk_definition}
\end{eqnarray}
that satisfy $u_{\bf k}v_{\bf k}=\Delta_{\bf k}/2E_{\bf k}=d_{1}/2d$. The sign of the gap ${\rm Sgn}(\Delta_{\bf k})={\rm Sgn}(d_{1})$ is unimportant in practice for $s$-wave SCs, but will be important for $d$-wave SCs. This is because when taking the derivative of momentum on $v_{\bf k}$, one only takes derivative on the square root but not on the sign
\begin{eqnarray}
\partial_{\mu}v_{\bf k}={\rm Sgn}(\Delta_{\bf k})\partial_{\mu}\sqrt{\frac{1}{2}\left(1-\frac{d_{3}}{d}\right)},
\label{vk_derivative}
\end{eqnarray}
because $v_{\bf k}$ and an infinitely small shift along ${\hat{\boldsymbol\mu}}$ direction $v_{{\bf k}+\delta k{\hat{\boldsymbol\mu}}}$ have the same sign if $\delta k\rightarrow 0$. The derivative $\partial_{\mu}v_{\bf k}$ is ill-defined at where the gap changes sign in a $d$-wave SC, rendering the quantum metric ill-defined along the nodal lines, as we shall see below.

We are interested in the quantum metric\cite{Provost80} $g_{\mu\nu}({\bf k})$ of the filled quasihole state $|n({\bf k})\rangle$ defined from the inner product of this state at momentum ${\bf k}$ and at momentum ${\bf k+\delta k}$
\begin{eqnarray}
|\langle n({\bf k})|n({\bf k+\delta k})\rangle|=1-\frac{1}{2}g_{\mu\nu}\delta k^{\mu}\delta k^{\nu},
\label{gmunu_definition}
\end{eqnarray}
which amounts to several equivalent expressions
\begin{eqnarray}
&&g_{\mu\nu}=\frac{1}{2}\langle\partial_{\mu}n|m\rangle\langle m|\partial_{\nu}\rangle+\left(\mu\leftrightarrow\nu\right)
\nonumber \\
&&=\frac{1}{4}\partial_{\mu}{\bf n}\cdot\partial_{\nu}{\bf n}=\left(u_{\bf k}\partial_{\mu}v_{\bf k}-v_{\bf k}\partial_{\mu}u_{\bf k}\right)\left(u_{\bf k}\partial_{\nu}v_{\bf k}-v_{\bf k}\partial_{\nu}u_{\bf k}\right)
\nonumber \\
&&=\frac{1}{4d^{4}}\left(d_{3}\partial_{\mu}d_{1}-d_{1}\partial_{\mu}d_{3}\right)
\left(d_{3}\partial_{\nu}d_{1}-d_{1}\partial_{\nu}d_{3}\right),
\label{bare_quantum_metric}
\end{eqnarray}
where $\partial_{\mu}\equiv\partial/\partial k^{\mu}$, and we have used Eqs.~(\ref{uk_vk_definition}) and (\ref{vk_derivative}). We see that either the derivative on the unit vector $\partial_{\mu}{\bf n}/2$ or on the Bogoliubov coefficients $\pm\left(u\partial_{\mu}v-v\partial_{\mu}u\right)$ can play the role of the vielbein, and the expression in terms of Bogoliubov coefficients in the second line of Eq.~(\ref{bare_quantum_metric}) is unique to SCs that has no analogy in semiconductors or insulators. Equation (\ref{bare_quantum_metric}) also implies that the volume form of the curved momentum space vanishes $\sqrt{\det g}=0$ for $D>1$, and consequently many geometrical quantities that involves integration over the curved momentum space would vanish at $D>1$, such as the Hilbert action $\int d^{D}{\bf k}\,\sqrt{\det g}R=0$ given by the integration of Ricci scalar $R$.

There is a very intuitive way to visualize the quantum metric using Bogoliubov coefficients. Suppose from the formula of quasihole state in Eq.~(\ref{quasihole_quasiparticle_states}), one writes the Bogoliubov coefficients into a two-component unit vector field ${\bf w}_{\bf k}=(v_{\bf k},-u_{\bf k})$ defined in a $D$-dimensional ${\bf k}$-space, which can also be regarded as representing the quasihole state as a unit vector in the Hilbert space. Then Eq.~(\ref{gmunu_definition}) can be rewritten as the dot product between the neighboring vectors
\begin{eqnarray}
&&\frac{1}{2}g_{\mu\nu}\delta k^{\mu}\delta k^{\nu}=1-|\langle n({\bf k})|n({\bf k+\delta k})\rangle|
\nonumber \\
&&=1-|{\bf w}_{\bf k}\cdot {\bf w}_{\bf k+\delta k}|.
\label{gmunu_vector_field}
\end{eqnarray}
Physically, this means that $g_{\mu\nu}$ can be simply understood as how much the product $|{\bf w}_{\bf k}\cdot {\bf w}_{\bf k+\delta k}|$ deviates from unity, which is equivalently how much the unit vector in the Hilbert space ${\bf w}_{\bf k}$ "twists" as one goes from ${\bf k}$ to ${\bf k+\delta k}$. If the ${\bf w}_{\bf k}$ is very uniform around ${\bf k}$, then $g_{\mu\nu}$ is small. In contrast, if ${\bf w}_{\bf k}$ changes its direction very dramatically around ${\bf k}$, meaning that $u_{\bf k}$ and $v_{\bf k}$ vary significantly near ${\bf k}$, then $g_{\mu\nu}$ is large. We will demonstrate this intuitive picture using concrete examples in the following sections.

Another geometrical quantity that we are interested is the momentum-integration of quantum metric
\begin{eqnarray}
&&{\cal G}_{\mu\nu}=\int\frac{d^{D}{\bf k}}{(2\pi)^{D}}g_{\mu\nu}({\bf k}),
\label{Gmunu_definition}
\end{eqnarray}
which we call the fidelity number\cite{deSousa23_fidelity_marker} (not to be confused with the fidelity of neighboring quasihole states $|\langle n({\bf k})|n({\bf k+\delta k})\rangle|$).Physically, this quantity represents the {\it average} distance between neighboring quasihole states $|n({\bf k})\rangle$ and $|n({\bf k+\delta k})\rangle$, and hence serves as a characteristic quantum geometrical property of the BZ torus. Moreover, it is also shown to be equivalent to the gauge-invariant part of the spread of Wannier functions\cite{Souza00,Marzari97,Marzari12} (in our case the Wannier function of the quasihole state). This quantity can be mapped to lattice sites in real space as a fidelity marker by considering a lattice Bogoliubov-de Gennes (BdG) Hamiltonian that has been diagonalized $H|E_{\ell}\rangle=E_{\ell}|E_{\ell}\rangle$. Introducing the projectors to the filled $E_{n}<0$ and empty $E_{m}>0$ lattice eigenstates from the projectors to the quasihole and quasiparticle states integrated over momentum
\begin{eqnarray}
&&{\hat P}=\sum_{n}\int\frac{d^{D}{\bf k}}{(2\pi)^{D}}|\psi_{n{\bf k}}\rangle\langle\psi_{n{\bf k}}|\rightarrow\sum_{n}|E_{n}\rangle\langle E_{n}|,
\nonumber \\
&&{\hat Q}=\sum_{m}\int\frac{d^{D}{\bf k'}}{(2\pi)^{D}}|\psi_{m{\bf k'}}\rangle\langle\psi_{m{\bf k'}}|\rightarrow \sum_{m}|E_{m}\rangle\langle E_{m}|,
\nonumber \\
\label{projector_PQ_k}
\end{eqnarray}
where $\langle{\bf r}|\psi_{n{\bf k}}\rangle=e^{i{\bf k\cdot r}/\hbar}\langle{\bf r}|n({\bf k})\rangle$ is the full quasihole state wave function (and likewisely for $|m({\bf k})\rangle$), it is found that the fidelity number can be written as
\begin{eqnarray}
&&{\cal G}_{\mu\nu}=\frac{1}{2}{\rm Tr}\left[{\hat P}\,{\hat r}_{\mu}{\hat Q}\,{\hat r}_{\nu}{\hat P}+{\hat P}\,{\hat r}_{\nu}{\hat Q}\,{\hat r}_{\mu}{\hat P}\right]
\nonumber \\
&&=\sum_{\bf r}{\cal G}_{\mu\nu}({\bf r}),
\label{totla_fidelity_definition}
\end{eqnarray}
where ${\hat r}_{\mu}$ and ${\hat r}_{\nu}$ are position operators on the lattice. The trace in this expression is over the lattice sites ${\bf r}$ and all the internal degrees of freedom $\alpha$ on every site. Treating the trace as a summation over all lattice sites ${\bf r}$, each term in the summation defines what we call the fidelity marker at site ${\bf r}$
\begin{eqnarray}
&&{\cal G}_{\mu\nu}({\bf r})=\frac{1}{2}\sum_{\alpha}\langle{\bf r,\alpha}|\left[{\hat P}\,{\hat r}_{\mu}{\hat Q}\,{\hat r}_{\nu}{\hat P}+{\hat P}\,{\hat r}_{\nu}{\hat Q}\,{\hat r}_{\mu}{\hat P}\right]|{\bf r,\alpha}\rangle
\nonumber \\
&&\equiv \frac{1}{2}\langle{\bf r}|\left[{\hat P}\,{\hat r}_{\mu}{\hat Q}\,{\hat r}_{\nu}{\hat P}+{\hat P}\,{\hat r}_{\nu}{\hat Q}\,{\hat r}_{\mu}{\hat P}\right]|{\bf r}\rangle,
\label{fidelity_marker_definition}
\end{eqnarray}
where the summation over $\alpha$ stands for summing over the spin up particle and spin down hole at site ${\bf r}$. The operator ${\hat {\cal G}_{\mu\nu}}\equiv\left[{\hat P}\,{\hat r}_{\mu}{\hat Q}\,{\hat r}_{\nu}{\hat P}+{\hat P}\,{\hat r}_{\nu}{\hat Q}\,{\hat r}_{\mu}{\hat P}\right]/2$ has been called fidelity operator. In the following sections, we shall see how the fidelity marker can be used to characterize the influence of real space inhomogeneity on the quantum geometry. 

%In the following sections, we shall see how the fidelity number and marker appear in various real space responses of SCs. 

\subsection{Electromagnetic responses of singlet SCs}

For a number of responses against external perturbations, such as a modulating scalar potential or electromagnetic wave, one often encounters the calculation of polarization operator. For singlet SCs, the polarization operator takes the general form\cite{Mahan00}
\begin{eqnarray}
&&P({\bf k,q},i\omega)
\nonumber \\
&&=-\sum_{\sigma\sigma '}\int_{0}^{\beta}d\tau\,e^{i\omega\tau}\langle T_{\tau}c_{\bf k+q\sigma}^{\dag}(\tau)c_{\bf k\sigma}(\tau)c_{\bf k'-q\sigma '}^{\dag}(0)c_{\bf k'\sigma '}(0)\rangle
\nonumber \\
&&=\frac{2}{\beta}\sum_{ip}\left[G({\bf k},ip)G({\bf k+q},ip+i\omega)\right.
\nonumber \\
&&\left.+F({\bf k},ip)F^{\dag}({\bf k+q},ip+i\omega)\right],
\end{eqnarray}
where ${\bf q}$ and ${\bf k}$ are external and internal momenta, respectively, $\sigma$ is the spin index, and $i\omega$ and $ip$ are Matsubara frequencies. The Green's functions are given by
\begin{eqnarray}
&&G({\bf k},ip)=\frac{u_{\bf k}^{2}}{ip-E_{\bf k}}+\frac{v_{\bf k}^{2}}{ip+E_{\bf k}},
\nonumber \\
&&F({\bf k},ip)=F^{\dag}({\bf k},ip)
\nonumber \\
&&=-u_{\bf k}v_{\bf k}\left(\frac{1}{ip-E_{\bf k}}-\frac{1}{ip+E_{\bf k}}\right).
\end{eqnarray}
We are interested in the retarded response at zero temperature, which is given by
\begin{eqnarray}
P_{0}({\bf k,q,}\,\omega)&=&2\left[\frac{u_{\bf k+q}^{2}v_{\bf k}^{2}-u_{\bf k}v_{\bf k}u_{\bf k+q}v_{\bf k+q}}{\hbar\omega-E_{\bf k}-E_{\bf k+q}+i\eta}\right.
\nonumber \\
&&\left.-\frac{v_{\bf k+q}^{2}u_{\bf k}^{2}-u_{\bf k}v_{\bf k}u_{\bf k+q}v_{\bf k+q}}{\hbar\omega+E_{\bf k}+E_{\bf k+q}+i\eta}\right].
\label{Pkqiw_ukvk}
\end{eqnarray}
after an analytical continuation $i\omega\rightarrow\hbar\omega+i\eta$, with $\eta$ a small artificial broadening. Furthermore, there are two kinds of situations that one often encounters in practical applications. One is the optical absorption process that corresponds to excitation of quasiparticles, which corresponds to taking the imaginary part of the first term in Eq.~(\ref{Pkqiw_ukvk}) at finite frequency, yielding 
\begin{eqnarray}
-\frac{1}{\pi}{\rm Im}P_{0}({\bf k,q,}\,\omega)&=&2\left[u_{\bf k+q}^{2}v_{\bf k}^{2}-u_{\bf k}v_{\bf k}u_{\bf k+q}v_{\bf k+q}\right]
\nonumber \\
&&\times\delta(\hbar\omega-E_{\bf k}-E_{\bf k+q}),
\label{ImP0}
\end{eqnarray}
where the $\delta$-function ensures the energy and momentum conservation. The other typical response is the static response that corresponds to taking the real part of both terms in Eq.~(\ref{Pkqiw_ukvk}) in the $\omega=0$ limit, yielding 
\begin{eqnarray}
{\rm Re}P_{0}({\bf k,q,}\,0)=-\frac{2(u_{\bf k+q}v_{\bf k}-v_{\bf k+q}u_{\bf k})^{2}}{E_{\bf k}+E_{\bf k+q}}.
\label{ReP0}
\end{eqnarray}
In what follows, we shall see some practical applications of these two situations, which turn out to both contain the integration of quantum metric. In particular, we will focus on the dynamic current-current correlator that is relevant to the infrared absorption, the static current-current correlator relevant to the paramagnetic current, and the static density-density correlator that is related to the dielectric function, and elaborate how quantum metric manifests in these quantities.

\subsubsection{Dynamic current-current correlator: Infrared absorption}

Consider a singlet SC subject to a transverse EM wave polarized in ${\hat{\boldsymbol\mu}}$ direction and propagating along ${\hat{\boldsymbol\nu}}$ direction with a small but finite wave vector ${\bf q}=q{\hat{\boldsymbol\nu}}$, so ${\hat{\boldsymbol\mu}}\perp{\hat{\boldsymbol\nu}}$. In this situation, the current density operator in $D$-dimension flowing along ${\hat{\boldsymbol\mu}}$ direction Fourier transformed along the propogation direction ${\hat{\boldsymbol\nu}}$ of the EM wave is
\begin{eqnarray}
&&j_{\mu}({\bf q})=\frac{e}{a^{D}}\sum_{\bf k}v_{\mu}({\bf k})c_{\bf k+q\sigma}^{\dag}c_{\bf k\sigma}.
\end{eqnarray}
where $v_{\mu}({\bf k})=\partial_{\mu}\varepsilon_{\bf k}$ is the normal state group velocity at ${\bf k}$ (not to be confused with the Bogoliubov coefficient $v_{\bf k}$). The perturbation is described by 
\begin{eqnarray}
H'=-a^{D}j_{\mu}({\bf q})A_{\mu}({\bf q},t),
\label{Hprime_jA}
\end{eqnarray}
where $A_{\mu}({\bf q},t)=\sum_{\bf r}A_{\mu}({\bf r},t)e^{i{\bf q\cdot r}}$ is the Fourier component of the time-dependent vector field polarized along ${\hat{\boldsymbol\mu}}$. Defining the Matsubara current-current correlator by
\begin{eqnarray}
&&\pi({\bf q},i\omega)=-\frac{a^{D}}{\hbar}\int_{0}^{\beta}d\tau\,e^{i\omega\tau}\langle T_{\tau}j_{\mu}({\bf q},\tau)j_{\mu}({\bf -q},0)\rangle
\nonumber \\
&&=\frac{e^{2}}{a^{D}}\sum_{\bf k}v_{\mu}^{2}P({\bf k,q},i\omega),
\label{current_current_correlator}
\end{eqnarray}
the optical conductivity $\sigma_{\mu\mu}({\bf q},\omega)\equiv\sigma({\bf q},\omega)$ along the direction of polarization ${\hat{\boldsymbol\mu}}$ for a clean SC at zero temperature can be calculated. According to the linear response theory, the conductivity corresponds to taking the imaginary part of the first term of Eq.~(\ref{Pkqiw_ukvk}) and correspondingly in Eq.~(\ref{current_current_correlator}) that represents the absorption process, and then integrated over momentum\cite{Mattis58,Mahan00}
\begin{eqnarray}
&&\sigma({\bf q},\omega)=\frac{2\pi e^{2}}{\omega}\int\frac{d^{D}{\bf k}}{(2\pi\hbar)^{D}}v_{\mu}^{2}
\nonumber \\
&&\times\left[u_{\bf k+q}^{2}v_{\bf k}^{2}-u_{\bf k}v_{\bf k}u_{\bf k+q}v_{\bf k+q}\right]\delta(\hbar\omega-E_{\bf k}-E_{\bf k+q}).
\nonumber \\
\label{optical_conductivity}
\end{eqnarray}
We shall see below how the coherence factor in this expression should be treated.

%since the factor in the integrand is much smaller than the unity $mv_{\mu}^{2}\Delta^{2}/E_{\bf k}^{2}(E_{\bf k}+\varepsilon_{\bf k})\ll 1$. Thus we see that the leading order optical conductivity at zero temperature is essentially $q^{2}$ times the fidelity marker spectral function, at least in this example where the normal state has a free electron dispersion. 

%{\cblue (2) Maybe generalize it to finite temperature. See Mahan's book how the Fermi distribution functions enter. }

\subsubsection{Static current-current correlator: Paramagnetic current}

In the presence of a static vector field that modulates with a finite wave vector ${\bf q}=q{\hat{\boldsymbol\nu}}$, the zero temperature London equation (in SI unit) is modified by\cite{Tinkham04}
\begin{eqnarray}
&&J_{\mu}({\bf q})=J_{\mu 1}({\bf q})+J_{\mu 2}({\bf q})=K_{1}({\bf q})A_{\mu}({\bf q})+J_{\mu 2}({\bf q}),\;\;\;
\label{J1J2_K}
\end{eqnarray}
where $J_{\mu 2}({\bf q})$ is the usual diamagnetic current that gives the Meissner effect, and is determined by the penetration depth $\lambda_{L}$ in 3D. Equation (\ref{J1J2_K}) also implies that we work in the London gauge where the vector potential is proportional to the current. The $J_{\mu 1}({\bf q})$ is a paramagnetic current that acts against $J_{\mu 2}({\bf q})$ and only occurs at ${\bf q}\neq{\bf 0}$. The paramagnetic current may be regarded as a response to the static vector field, described by the perturbation in Eq.~(\ref{Hprime_jA}) but with a static $A_{\mu}({\bf q},t)=A_{\mu}({\bf q})$, and be cauculated by a linear response theory
\begin{eqnarray}
J_{\mu 1}({\bf q})=\langle j_{\mu}({\bf q})\rangle=-{\rm Re}\,\pi({\bf q},0)|_{T=0}A_{\mu}({\bf q}),
\end{eqnarray}
yielding the response coefficient 
\begin{eqnarray}
&&K_{1}({\bf q})={\rm Re}\,\pi({\bf q},0)|_{T=0}
=\frac{e^{2}}{a^{D}}\sum_{\bf k}v_{\mu}^{2}{\rm Re}P_{0}({\bf k,q},0)
\nonumber \\
&&=-2e^{2}\int\frac{d^{D}{\bf k}}{(2\pi\hbar)^{D}}\,v_{\mu}^{2}\,\frac{(u_{\bf k+q}v_{\bf k}-v_{\bf k+q}u_{\bf k})^{2}}{E_{\bf k}+E_{\bf k+q}},
\label{K1_ukvk}
\end{eqnarray}
which agrees with the result directly calculated from applying first order perturbation theory to the BCS ground state\cite{Tinkham04}.

\subsubsection{Static density-density correlator: Linear screening}

The static $\omega=0$ dielectric function at zero temperature within random phase approximation (RPA) is given by\cite{Mahan00} 
\begin{eqnarray}
\varepsilon({\bf q},\,0)=1-V_{\bf q}P_{0}({\bf q},\,0),
\end{eqnarray}
where $V({\bf q})=\sum_{\bf r}e^{i{\bf q\cdot r}}V({\bf r})$ is the Fourier transform of the Coulomb potential, and $P_{0}({\bf q},\,0)$ is precisely that in Eq.~(\ref{ReP0}) integrated over the internal momentum ${\bf k}$
\begin{eqnarray}
&&P_{0}({\bf q},\,0)=\int\frac{d^{D}{\bf k}}{(2\pi\hbar/a)^{D}}{\rm Re}P_{0}({\bf k,q},\,0)
\nonumber \\
&&=-2\int\frac{d^{D}{\bf k}}{(2\pi\hbar/a)^{D}}\frac{(u_{\bf k+q}v_{\bf k}-v_{\bf k+q}u_{\bf k})^{2}}{E_{\bf k}+E_{\bf k+q}}.
\label{dielectric_P0}
\end{eqnarray}
The result is an expression very similar to that in Eq.~(\ref{K1_ukvk}).

\section{s-wave superconductors}

\subsection{Mean field theory for s-wave SCs}

Our first concrete example concerns the clean, prototype s-wave superconductor in which analytical results for quantum metric can be given. For simplicity, we consider the $D$-dimensional cubic lattice models with nearest-neighbor hopping $t$ and chemical potential $\mu$, in which the dispersion is given by
\begin{eqnarray}
\varepsilon_{\bf k}=-2t\sum_{i=1}^{D}\cos k_{i}-\mu,
\label{swave_tight_binding}
\end{eqnarray}
The gap is a constant $d_{1}=\Delta$ and is treated as a parameter, hence $\partial_{\mu}d_{1}=0$, and the derivative on the normal state dispersion $\partial_{\mu}d_{3}=\partial_{\mu}\varepsilon_{\bf k}=v_{\mu}({\bf k})\equiv v_{\mu}$ just gives the normal state group velocity along $\mu$-direction at momentum ${\bf k}$. Using Eq.~(\ref{bare_quantum_metric}), the quantum metric is 
\begin{eqnarray}
g_{\mu\nu}=\frac{\Delta^{2}v_{\mu}v_{\nu}}{4E_{\bf k}^{4}}.
\label{zero_T_quantum_metric}
\end{eqnarray}
We see that $\Delta v_{\mu}/2E_{\bf k}$ plays the role of vielbein.
We are particularly interested in the region near the Fermi momentum $k\approx k_{F}$ where the dispersion can be expanded by $\varepsilon_{\bf k}\approx v_{F}(k-k_{F})$. In addition, the BCS coherence length in the mean field theory is given by
\begin{eqnarray}
\xi=\frac{\hbar v_{F}}{\pi\Delta},
\end{eqnarray}
Denoting the quantum metric exactly at the Fermi momentum as $g_{\mu\nu}({\bf k}_{F})$, the quantum metric near the Fermi momentum takes the Lorentzian form
\begin{eqnarray}
&&g_{\mu\nu}({\bf k\approx k}_{F})\approx\frac{g_{\mu\nu}({\bf k}_{F})}{1+2\pi^{2}(\xi/\hbar)^{2}(k-k_{F})^{2}},
\nonumber \\
&&g_{\mu\nu}({\bf k}_{F})=\frac{v_{\mu}v_{\nu}}{4\Delta^{2}}.
\label{gmunu_Lorentzian}
\end{eqnarray}
This simple formula allows us to plot the profile of the quantum metric in momentum space.

To apply the optical conductivity formula in Eq.~(\ref{optical_conductivity}) to $s$-wave SCs, we expand the coherence factor to second order in ${\bf q}=q{\hat{\boldsymbol\nu}}$, yielding 
\begin{eqnarray}
&&u_{\bf k+q}^{2}v_{\bf k}^{2}-u_{\bf k}v_{\bf k}u_{\bf k+q}v_{\bf k+q}
\nonumber \\
&&\approx(qu_{\bf k}v_{\bf k}+q^{2}v_{\bf k}\partial_{\nu}u_{\bf k})(v_{\bf k}\partial_{\nu}u_{\bf k}-u_{\bf k}\partial_{\nu}v_{\bf k})
\nonumber \\
&&=\left(\frac{q\Delta}{2E_{\bf k}}+\frac{q^{2}\Delta^{3}v_{\nu}}{4(E_{\bf k}+\varepsilon_{\bf k})E_{\bf k}^{3}}\right)\frac{\Delta v_{\nu}}{2E_{\bf k}^{2}}.
\end{eqnarray}
Moreover, we will approximate the argument in the $\delta$-function by $\hbar\omega-E_{\bf k}-E_{\bf k+q}\approx \hbar\omega-2E_{\bf k}$, which allows to replace the inverse frequency in the expression by $1/\omega\approx \hbar/2E_{\bf k}$. These approximations are justified because, given the typical band gap $\Delta\sim 0.01$eV of $s$-wave SCs, the minimal wave vector of the light that can excite quasiparticles is much smaller than the Fermi momentum $q\ll k_{F}$, allowing us to expand the optical conductivity to first and second order in $q$. It should be reminded that this expansion in $q$ is appropriate for clean SCs where momentum conservation is satisfied, in contrast to the seminal work of Mattis and Bardeen that discuss the limit of dirty SCs where ${\bf k}$ and ${\bf k+q}$ are treated as two unrelated momenta\cite{Mattis58,Mahan00}.

To apply the paramagnetic current formula in Eq.~(\ref{K1_ukvk}) to $s$-wave SC, we observe that the expansion of the coherence factor to leading order in ${\bf q}=q{\hat{\boldsymbol\nu}}$ yields the diagonal element of the quantum metric along ${\bf q}$
\begin{eqnarray}
&&(u_{\bf k+q}v_{\bf k}-v_{\bf k+q}u_{\bf k})^{2}
\nonumber \\
&&\approx q^{2}\left(v_{\bf k}\partial_{\nu}u_{\bf k}-u_{\bf k}\partial_{\nu}v_{\bf k}\right)^{2}=q^{2}g_{\nu\nu},
\label{swave_udv_vdk_expansion}
\end{eqnarray}
according to Eq.~(\ref{bare_quantum_metric}). In addition, an analytical expression for these electromagnetic responses can be given for continuous models with a quadratic dispersion, which has the expressions of energy dispersion, quantum metric, and coherence length near the Fermi surface
\begin{eqnarray}
&&E_{\bf k}=\left[\left(\frac{k^{2}}{2m}-\frac{k_{F}^{2}}{2m}\right)^{2}+\Delta^{2}\right]^{1/2}
\nonumber \\
&&\approx\Delta\left[1+\frac{1}{2}\left(\frac{\pi\xi}{\hbar}\right)^{2}(k-k_{F})^{2}\right],
\nonumber \\
&&g_{\nu\nu}\approx\frac{(k\cos\theta/m)^{2}}{4\Delta^{2}\left[1+2\left(\frac{\pi\xi}{\hbar}\right)^{2}(k-k_{F})^{2}\right]},
\nonumber \\
&&\xi=\hbar k_{F}/\pi m\Delta.
\label{Ek_gnunu_approx}
\end{eqnarray} 
Moreover, because the width of Lorentzians is extremely small $1/\xi\ll k_{F}$, one may approximate them as $\delta$-functions
\begin{eqnarray}
&&\frac{1}{\left[1+\frac{1}{2}\left(\frac{\pi\xi}{\hbar}\right)^{2}(k-k_{F})^{2}\right]^{n}}
\nonumber \\
&&\approx\frac{1}{\left[1+2n\pi^{4}\left(\frac{\xi}{a}\right)^{2}(x-x_{F})^{2}\right]}
\nonumber \\
&&=\frac{\eta^{2}}{\eta^{2}+(x-x_{F})^{2}}\approx\pi\eta\delta(x-x_{F})
\label{Lorentzian_deltafn}
\end{eqnarray}
after a change of variable $x=k/(2\pi\hbar/a)$ and defining $\eta=a/\sqrt{2n}\pi^{2}\xi$, where $n$ is any power in the calculation. These approximations allow to express the electromagnetic responses in terms of the fidelity number, as we shall see below for the 3D and 2D cases.

%We see that as the gap vanishes $\Delta\rightarrow 0$ near the critical temperature $T\rightarrow T_{c}$, the quantum metric at the Fermi momentum $g_{\mu\nu}({\bf k}_{F})$ diverges like $\Delta^{-2}$, and the Lorentzian peak of $g_{\mu\nu}({\bf k\approx k}_{F})$ gradually narrows because of the increasing $\xi$.

%{\cblue (1) I think the quantum metric of cuprates may help to clarify whether cuprates are really close to a quantum critical point at some doping.}

\subsection{Quantum geometry and electromagnetic response of 3D $s$-wave SCs}

\subsubsection{Profile of the quantum metric in momentum space}

\begin{figure}[ht]
\begin{center}
\includegraphics[clip=true,width=0.99\columnwidth]{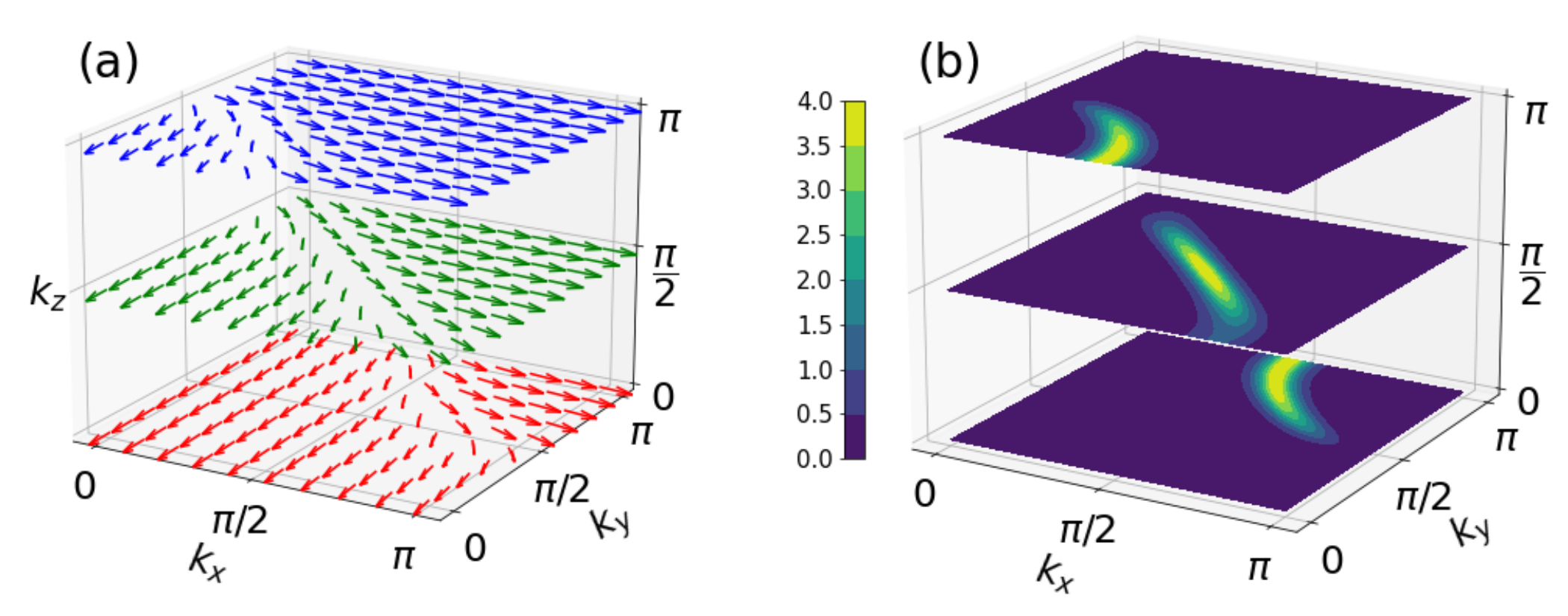}
\caption{(a) The Bogoliubov coefficients plotted as a two-component vector field $(v_{\bf k},-u_{\bf k})$ in the momentum space of a 3D $s$-wave SC. Without loss of generality the two-component vectors are chosen to be lying on the $xy$-plane. (b) The magnitude of quantum metric $g_{xx}$ of the 3D $s$-wave SC in momentum space, which coincides with the twisting of the vector field $(v_{\bf k},-u_{\bf k})$ in (a).} 
\label{fig:3DSC_results}
\end{center}
\end{figure}

To simulate the $s$-wave SC on a 3D cubic lattice, we use the tight-binding model in Eq.~(\ref{swave_tight_binding}) with $t=1$ and $\mu=-0.2$, and a rather large gap $\Delta=0.5$ just to visualize the effect. The profile of the Bogoliubov coefficients $(v_{\bf k},-u_{\bf k})$ plotted as a two-component vector field in the 3D momentum space, which can also be considered as the quasihole state $|n\rangle$ as a unit vector in the Hilbert space at momentum ${\bf k}$, is shown in Fig.~\ref{fig:3DSC_results} (a). One sees that both below and above the Fermi surface, the vector field is fairly uniform, indicating that the wave function is either hole-like or electron-like. Only near the Fermi surface does the vector field start to twist dramatically in order to go from hole-like to electron-like. As a result, the quantum metric $g_{xx}({\bf k})$ peaks at the Fermi surface of the normal state, in accordance with Eq.~(\ref{gmunu_vector_field}).

%Interestingly, we find that the magnitude of $g_{xx}^{d}$ at different values of $k_{z}$ plotted as a function of $(k_{x},k_{y})$ does not change much, but the magnitude of $g_{xx}^{d}$ at different values of $k_{x}$ plotted as a function of $(k_{y},k_{z})$ changes very much since the Fermi velocity $v_{x}=2t\sin k_{x}$ changes with $k_{x}$. Because $g_{xx}^{d}$ peaks at the Fermi surface of the normal state, the fidelity number spectral function ${\cal G}_{xx}^{d}(\omega)$ peaks at the band gap $\hbar\omega\approx 2\Delta$, since it is the states near the Fermi surface that are participating the pairing and contribute the most quantum metric.

%{\cblue (1) Maybe there is an analytical expression for the fidelity number spectral function too? }

For the continuous model, by using the approximations in Eqs.~(\ref{Ek_gnunu_approx}) and (\ref{Lorentzian_deltafn}), one can obtain the analytical expression for the fidelity number
\begin{eqnarray}
&&{\cal G}_{\mu\mu}^{3D}=\int\frac{d^{3}{\bf k}}{(2\pi)^{3}}g_{\mu\mu}
\nonumber \\
&&=\frac{\pi^{2}}{6\sqrt{2}}\left(\frac{\xi}{a}\right)\left(\frac{k_{F}}{2\pi\hbar/a}\right)^{2}\left(\frac{\hbar}{a}\right).
\label{Gmumu_3D_analytical}
\end{eqnarray}
The factor $k_{F}/(2\pi\hbar/a)$ in this expression is of the order of unity, so the fidelity number is essentially given by the coherence length divided by the lattice constant $\xi/a$ times the correct unit $\hbar/a$. Remarkably, this result implies that the fidelity number, or equivalently the spread of Wannier function\cite{Souza00,Marzari97,Marzari12}, is synonymous to the coherence length, and hence any property of SCs that is proportional to the coherence length can as well be written in terms of the fidelity number.

For the typical coherence length $\xi\sim\mu$m and lattice constant $a\sim$nm of 3D $s$-wave SCs\cite{Tinkham04}, one obtains ${\cal G}_{\mu\mu}^{3D}\sim 10^{3}(\hbar/a)$. This value can be compared with the fidelity number in 3D topological insulators (TIs) whose dimensionless part scales like $|M|a/\hbar v+{\rm const}\sim {\cal O}(1)$, where $v$ is the Fermi velocity and $M$ is the band gap, which yields a number that is of the order of unity\cite{deSousa23_fidelity_marker}. Thus we see that the fidelity number of $s$-wave SCs is actually 2 to 3 orders of magnitude larger than that of a typical TI, indicating that the BZ manifold of an $s$-wave SC is much more distorted, or equivalently the Wannier functions are much more spread out\cite{Souza00,Marzari97,Marzari12}, in comparison with that of an 3D TI.

\subsubsection{Infrared absorption}

For 3D $s$-wave SC, the optical conductivity in Eq.~(\ref{optical_conductivity}) expanded to first order in ${\bf q}=q{\hat{\boldsymbol\nu}}$ vanishes
\begin{eqnarray}
&&\sigma^{\rm 1st}({\bf q},\omega)\approx \pi e^{2}\hbar q\int\frac{d^{3}{\bf k}}{(2\pi\hbar)^{3}}g_{\mu\mu}v_{\nu}\delta(\hbar\omega-2E_{\bf k})
\nonumber \\
&&=0,
\end{eqnarray}
since the quantum metric is even but the velocity is odd in ${\bf k}$. Thus the first nonvanishing contribution is second order in $q$
\begin{eqnarray}
\sigma^{\rm 2nd}({\bf q},\omega)&\approx& 2\pi e^{2}\frac{\hbar q^{2}}{2m}\int\frac{d^{3}{\bf k}}{(2\pi\hbar)^{3}}\left[\frac{mv_{\nu}^{2}\Delta^{2}}{E_{\bf k}^{2}(E_{\bf k}+\varepsilon_{\bf k})}\right]
\nonumber \\
&&\times g_{\mu\mu}\delta(\hbar\omega-2E_{\bf k}),
\label{sigmaw_q2}
\end{eqnarray}
given by the integration of quantum metric $g_{\mu\mu}$ weighted by the dimensionless factor $mv_{\nu}^{2}\Delta^{2}/E_{\bf k}^{2}(E_{\bf k}+\varepsilon_{\bf k})$ and the energy conservation condition. This expression is conceptually different from the optical conductivity in semiconductors, where the quantum metric is exactly the matrix element for the excitation of electrons from the valence to the conduction band\cite{Ozawa18,vonGersdorff21_metric_curvature,Chen22_dressed_Berry_metric}. In contrast, the Bogoliubov transformation renders a more complicated form for the matrix element. Nevertheless, the matrix element in Eq.~(\ref{sigmaw_q2}) still contains the contribution from the quantum metric.

\subsubsection{Paramagnetic current}

For 3D $s$-wave SC, using Eqs.~(\ref{K1_ukvk}) and (\ref{swave_udv_vdk_expansion}) yields the response coefficient for the paramagnetic current
\begin{eqnarray}
K_{1}^{3D}({\bf q})\approx-e^{2}q^{2}\int\frac{d^{3}{\bf k}}{(2\pi\hbar)^{3}}v_{\mu}^{2}\frac{g_{\nu\nu}}{E_{\bf k}}.
\label{K1_swave_3D}
\end{eqnarray}
For the continuous model, we may define the ${\hat{\boldsymbol\nu}}$ direction to be along the solid angles $(\theta,\phi)$ that are to be integrated out, and the velocity factor to be $v_{\mu}=\cos(\theta-\alpha)k/m$, where $\alpha$ is the angle between the polarization ${\hat{\boldsymbol\mu}}$ and the spatial modulation ${\hat{\boldsymbol\nu}}\parallel{\hat{\bf q}}$ directions of the vector field ${\bf A}$. The integration in the spherical coordinate can then be carried out using Eqs.~(\ref{Ek_gnunu_approx}) and (\ref{Lorentzian_deltafn}), yielding
\begin{eqnarray}
&&K_{1}^{3D}({\bf q})\approx-e^{2}\frac{4\pi^{6}f(\alpha)}{\sqrt{10}ma^{3}}\left(\frac{\xi}{a}\right)^{2}
\left(\frac{q}{2\pi\hbar/a}\right)^{2}\left(\frac{k_{F}}{2\pi\hbar/a}\right)^{3},
\nonumber \\
&&f(\alpha)\equiv \frac{4}{15}+\frac{2}{15}\cos^{2}\alpha.
\label{K1q_3D_expression1}
\end{eqnarray}
The $k_{F}/(2\pi\hbar/a)$ is again of the order of unity, and $q/(2\pi\hbar/a)$ is the spatial modulation of vector field measured in unit of Fermi wavelength, and $e^{2}/ma^{3}$ is the correct unit for $K_{1}({\bf q})$ in 3D. Note that the factor $\left[k_{F}/(2\pi\hbar/a)\right]^{3}$ essentially represents the volume of the Fermi sea measured in units of the BZ, which also roughly represents the electron density. For the most situations, the polarization and propagation of the vector field are perpendicular ${\hat{\boldsymbol\mu}}\perp{\hat{\boldsymbol\nu}}$, yielding $\alpha=\pi/2$ and the angular factor is just $f(\alpha)=4/15$. Equation (\ref{K1q_3D_expression1}) means that $K_{1}({\bf q})$ is essentially given by the square of coherence length measured in units of lattice constant $(\xi/a)^{2}\sim 10^{6}$, which can reach a very large number. Moreover, it can be expressed in terms of the fidelity number ${\cal G}_{\nu\nu}$ in eq.~(\ref{Gmumu_3D_analytical}) as
\begin{eqnarray}
K_{1}^{3D}({\bf q})&=&-\frac{288\pi^{2}}{\sqrt{10}}\frac{f(\alpha)e^{2}}{ma^{3}}
\left(\frac{q}{2\pi\hbar/a}\right)^{2}
\nonumber \\
&&\times\left(\frac{k_{F}}{2\pi\hbar/a}\right)^{-1}\left(\frac{{\cal G}_{\nu\nu}^{3D}}{\hbar/a}\right)^{2},
\end{eqnarray}
manifesting a quadratic dependence on the fidelity number. Note that the quadratic dependence on $q$ is well-known in the literature\cite{Tinkham04}, and our calculation gives the prefactor of this dependence a quantum geometrical interpretation.

\subsubsection{Linear screening}

For the linear screening, putting the expansion in Eq.~(\ref{swave_udv_vdk_expansion}) into the dielectric response in Eq.~(\ref{dielectric_P0}) and approximate $E_{\bf k}+E_{\bf k+q}\approx 2E_{\bf k}$ yield
\begin{eqnarray}
&&P_{0}^{3D}({\bf q},0)\approx-q^{2}\int\frac{d^{3}{\bf k}}{(2\pi\hbar/a)^{3}}\,\frac{g_{\nu\nu}}{E_{\bf k}}.
\label{P0_swave_3D}
\end{eqnarray}
Using the approximations in Eqs.~(\ref{Ek_gnunu_approx}) and (\ref{Lorentzian_deltafn}), and also using the fidelity number in Eq.~(\ref{Gmumu_3D_analytical}), for the continuous model we obtain
\begin{eqnarray}
&&P_{0}^{3D}({\bf q},0)\approx-\frac{4\pi^{4}}{3\sqrt{10}\Delta}\left(\frac{\xi}{a}\right)
\left(\frac{k_{F}}{2\pi\hbar/a}\right)^{2}\left(\frac{q}{2\pi\hbar/a}\right)^{2}
\nonumber \\
&&=-\frac{8\pi^{2}}{\sqrt{5}\Delta}\left(\frac{q}{2\pi\hbar/a}\right)^{2}\frac{{\cal G}_{\nu\nu}^{3D}}{\hbar/a},
\end{eqnarray}
which may be used to extract ${\cal G}_{\nu\nu}$ provided that the lattice constant $a$ and gap $\Delta$ are known.

%{\cblue (1) Interesting, this means that the ratio $K_{1}({\bf q})/P_{0}({\bf q},0)$ is a universal number that does not depend on many details of the SC? And moreover this ratio will get rid of the $q^{2}$ factor. }

\begin{figure}[ht]
\begin{center}
\includegraphics[clip=true,width=0.99\columnwidth]{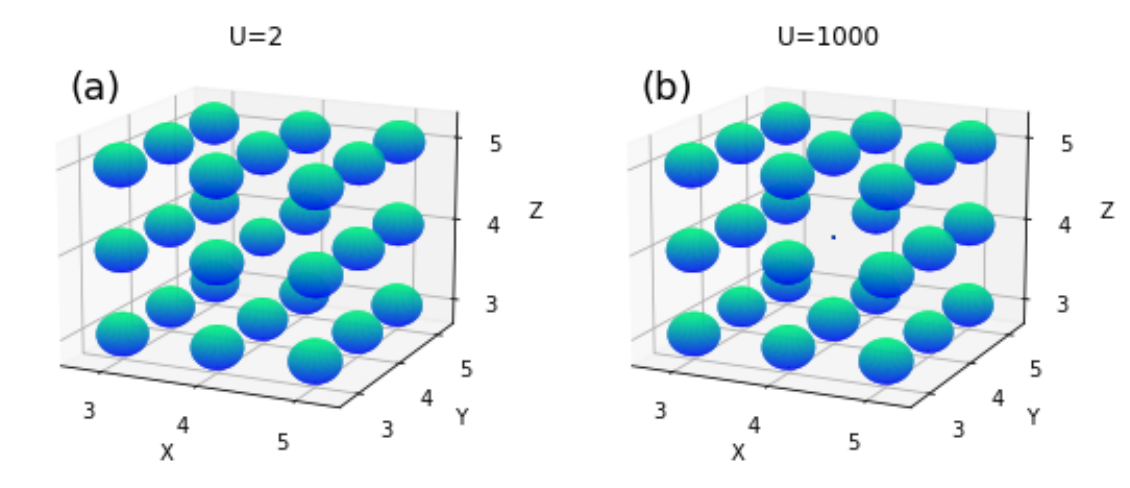}
\caption{The fidelity marker in a 3D $s$-wave SC around a nonmagnetic impurity with (a) a weak impurity potential $U=2$, where the largest sphere represents magnitude 0.332, and (b) a strong impurity potential $U=1000$, where the largest sphere corresponds to 0.339. } 
\label{fig:3DSC_impurity_marker}
\end{center}
\end{figure}

\subsubsection{Fidelity marker near an impurity \label{sec:fidelity_marker_3DSC}}

In Fig.~\ref{fig:3DSC_impurity_marker} (a) and (b), we show the central region of a cubic lattice in which we perform numerical calculation for the fidelity marker ${\cal G}_{xx}({\bf r})$ in the presence of a nonmagnetic impurity with local impurity potential $U$. The marker is fairly constant for sites far away from the impurity, but it is locally suppressed on the impurity site. Increasing the impurity potential further suppresses the marker until it is completely diminished, as can be seen by comparing Fig.~\ref{fig:3DSC_impurity_marker} (a) and (b). Through calculating the spatial average of the marker, we find that the average marker is suppressed by the impurity, indicating that nonmagnetic impurities reduce the average distance between the quasihole state in momentum space for a 3D $s$-wave SC.

\begin{figure}[ht]
\begin{center}
\includegraphics[clip=true,width=0.99\columnwidth]{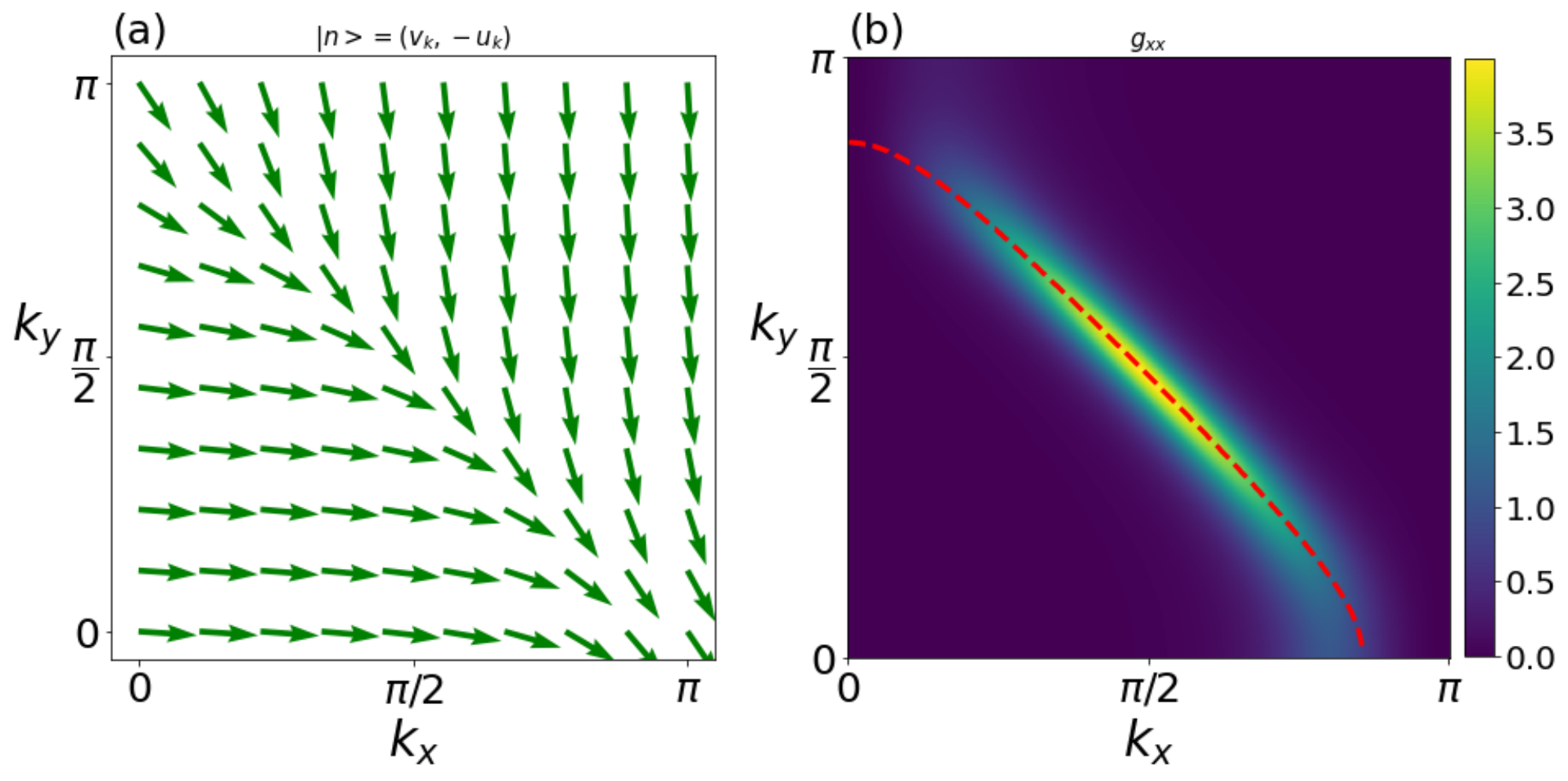}
\caption{(a) The Bogoliubov coefficients as a 2D vector field for 2D $s$-wave SC in the first quartet of the BZ. The twisting of this vector field under a small displacement $\delta k_{x}$ along ${\hat{\bf x}}$ direction is essentially the quantum metric $g_{xx}({\bf k})$ shown in (b), which peaks at the Fermi surface (dotted line).  } 
\label{fig:2DSC_results}
\end{center}
\end{figure}

\subsection{Quantum geometry and electromagnetic response of 2D $s$-wave SCs}

\subsubsection{Profile of the quantum metric in momentum space}

An SC may be considered 2D if its thickness is smaller than the in-plane coherence length\cite{Saito16}. We will consider strictly 2D systems with an $s$-wave pairing, and assume that the Mermin-Wigner theorem\cite{Mermin66} can be overcome by some other factors not included in the mean field theory, such as weak coupling between the planes. The Bogoliubov coefficients as a 2D vector field is shown in Fig.~\ref{fig:2DSC_results} (a). At a momentum ${\bf k}$, the twisting of this vector field under a small displacement $\delta k_{x}$ along ${\hat{\bf x}}$ direction gives the quantum metric $g_{xx}({\bf k})$ shown in Fig.~\ref{fig:2DSC_results}. We obtain a profile of $g_{xx}({\bf k})$ that highly peaks at the Fermi surface, in agreement with Eq.~(\ref{gmunu_vector_field}).

For the continuous model of 2D $s$-wave SCs, analytically carrying out the polar integration using the approximations in Eq.~(\ref{Ek_gnunu_approx}) and (\ref{Lorentzian_deltafn}) yields the fidelity number
\begin{eqnarray}
{\cal G}_{\mu\mu}^{2D}=\int\frac{d^{2}{\bf k}}{(2\pi)^{2}}g_{\mu\mu}
\approx\frac{\pi^{2}}{8\sqrt{2}}\left(\frac{\xi}{a}\right)\left(\frac{k_{F}}{2\pi\hbar/a}\right).
\label{Gmumu_2D_analytical}
\end{eqnarray}
Once again the $k_{F}/(2\pi\hbar/a)$ factor is of the order of unity, so we see that the fidelity number is a dimensionless number determined by the coherence length measured in units of lattice constant $\xi/a$, just like in the 3D case. As a result, any property that is proportional to the coherence length is a direct measurement of the fidelity number. Note that various 2D SCs with evidence for $s$-wave pairing have been discovered\cite{Cao15,Xi16,Onishi16,Ugeda16,Hotta16,Qiu21}, although not much information about their coherence length has been extracted. Nevertheless, within the BCS framework and estimating from their low critical temperatures, the coherence length of these materials should also be of the order of $\mu$m, yielding a fidelity number $\sim 10^{3}$. This number is much larger than that in 2D TIs which is logarithmic to the band gap $\sim\ln |M|a/\hbar v$ and hence of the order of unity, indicating a much more distorted BZ manifold in 2D $s$-wave SCs.

\subsubsection{Infrared absorption}

The infrared absorption in 2D is precisely that in 3D given by Eq.~(\ref{sigmaw_q2}) with a reduction of the dimension of integration $\int d^{3}{\bf k}/(2\pi\hbar)^{3}\rightarrow\int d^{2}{\bf k}/(2\pi\hbar)^{2}$. As a result, the quantum metric still enters the integrand of the momentum-integration.

\subsubsection{Paramagnetic current}

The paramagnetic current in 2D is given by that in Eq.~(\ref{K1_swave_3D}) with a reduction in the integration $\int d^{3}{\bf k}/(2\pi\hbar)^{3}\rightarrow\int d^{2}{\bf k}/(2\pi\hbar)^{2}$. The analytical result for the continuous model is 
\begin{eqnarray}
&&K_{1}^{2D}({\bf q})=-\frac{2\pi^{5}e^{2}\overline{f}(\alpha)}{\sqrt{10}ma^{2}}
\left(\frac{q}{2\pi\hbar/a}\right)^{2}
\left(\frac{\xi}{a}\right)^{2}\left(\frac{k_{F}}{2\pi\hbar/a}\right)^{2}
\nonumber \\
&&=-\frac{256\pi e^{2}\overline{f}(\alpha)}{\sqrt{10}ma^{2}}\left(\frac{q}{2\pi\hbar/a}\right)^{2}\left({\cal G}_{\nu\nu}^{2D}\right)^{2},
\nonumber \\
&&\overline{f}(\alpha)\equiv\frac{\pi}{4}+\frac{\pi}{2}\cos^{2}\alpha,
\end{eqnarray}
which is quadratic in the fidelity number.

\subsubsection{Linear screening}

The linear screening in 2D can be calculated from replacing $\int d^{3}{\bf k}/(2\pi\hbar/a)^{3}\rightarrow\int d^{2}{\bf k}/(2\pi\hbar/a)^{2}$ in Eq.~(\ref{P0_swave_3D}). Applying the approximations in Eqs.~(\ref{Ek_gnunu_approx}) and (\ref{Lorentzian_deltafn}) yields the result for the continuous model
\begin{eqnarray}
&&P_{0}^{2D}({\bf q},0)\approx-\frac{\pi^{4}}{\sqrt{10}\Delta}\left(\frac{\xi}{a}\right)\left(\frac{k_{F}}{2\pi\hbar/a}\right)
\left(\frac{q}{2\pi\hbar/a}\right)^{2}
\nonumber \\
&&=-\frac{8\pi^{2}}{\sqrt{5}\Delta}\left(\frac{q}{2\pi\hbar/a}\right)^{2}{\cal G}_{\nu\nu}^{2D}
\end{eqnarray}
which again implies that ${\cal G}_{\nu\nu}^{2D}$ can be measured by detecting $P_{0}^{2D}({\bf q},0)$ in 2D the $q\rightarrow 0$ limit.

%{\cblue (1) Mention the simple geometrical interpretation of a cylinder and a human arm in Appendix \ref{apx:cylinder_metric}. }

\subsubsection{Fidelity marker near an impurity}

\begin{figure}[ht]
\begin{center}
\includegraphics[clip=true,width=0.99\columnwidth]{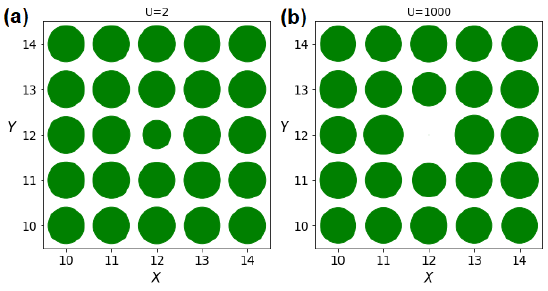}
\caption{The fidelity marker in a 2D $s$-wave SC around a nonmagnetic impurity with (a) impurity potential $U=2$, where the largest circle represents magnitude 0.433, and (b) $U=1000$, where the largest circle corresponds to 0.474. } 
\label{fig:2DSC_impurity_marker}
\end{center}
\end{figure}

The fidelity marker around a nonmagnetic impurity in a square lattice of 2D $s$-wave SC is shown in Fig.~\ref{fig:2DSC_impurity_marker} (a) and (b). We find a behavior similar to that of 3D $s$-wave shown in Sec.~\ref{sec:fidelity_marker_3DSC}, namely the marker is locally suppressed on the impurity site by the impurity potential, causing the average marker to be reduced. This implies that nonmagnetic impurities also reduce the average distance between the quasihole state in momentum space of a 2D $s$-wave SC.

%with a spatial profile that decays monotonically with the distance away from the impurity site. ON the other hand, for a magnetic impurity, our numerical result shown in Fig.~\ref{fig:2DSC_impurity_marker} (d) indicates that the marker is also suppressed on the impurity site, but has lobes along four diagonal directions where the magnitude of the marker is enhanced. 

%{\cblue (1) Actually what determines the decay length of the fidelity marker? It seems like it is not the coherence length because the coherence length is very long, but the marker decays much rapidly. I think we should plot the local density and local gap and see their profile in comparison with the marker. Also do different values of the gap and see how the profiles change. }

%To investigate the quantum geometric properties of the momentum space manifold $d$-wave SCs, our strategy is to solve the SC gap by means of a mean field theory directly on the lattice in real space, and then put the resulting gap into momentum space to find the quantum metric. For this purpose, we consider the lattice model of a $d$-wave SC
%\begin{eqnarray}
%H=-\sum_{ij\sigma}t_{ij}c_{i\sigma}^{\dag}c_{j\sigma}-\mu\sum_{i\sigma}c_{i\sigma}^{\dag}c_{i\sigma}
%-V\sum_{\langle ij\rangle}n_{i\uparrow}n_{j\downarrow},
%\end{eqnarray}
%where the nearest-neighbor attractive interaction $V$ induces the $d$-wave pairing. The order parameter $\Delta_{\delta i}=V\langle c_{i\uparrow}c_{i+\delta\downarrow}\rangle$ is obtained by solving the Bogoliubov-de Gennes equation {\cblue (Note that my 2009 paper uses spin-dependent BdG, but here I don't need this.)}

\begin{figure}[ht]
\begin{center}
\includegraphics[clip=true,width=0.99\columnwidth]{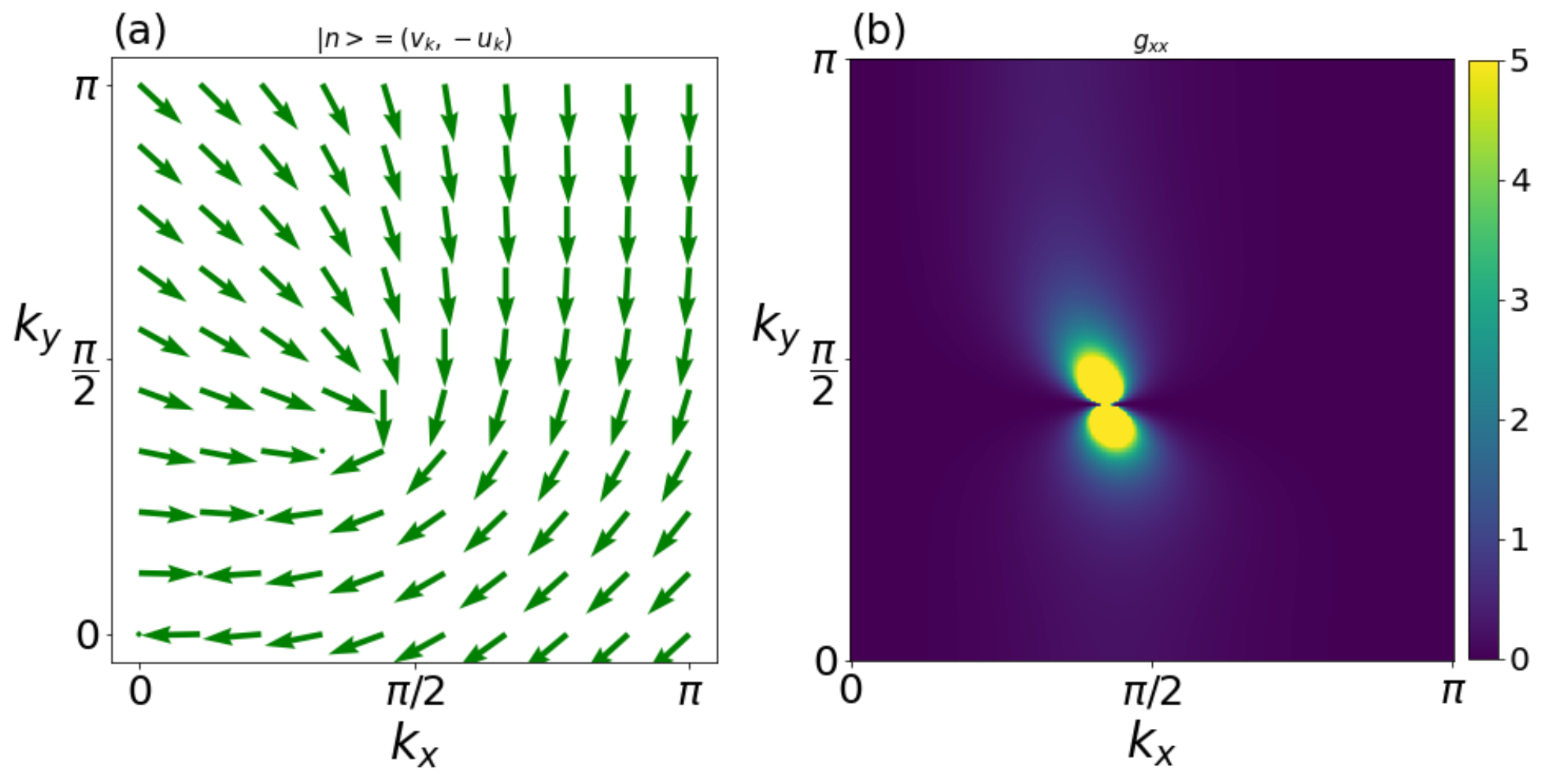}
\caption{(a) The vector field of $(v_{\bf k},-u_{\bf k})$ in the momentum space of a $d$-wave SC, and (b) the quantum metric $g_{xx}({\bf k})$ that corresponds to the twisting of this vector field. } 
\label{fig:2D_DSC_quantum_metric}
\end{center}
\end{figure}

\section{d-wave superconductors}

\subsection{Mean field theory for $d$-wave SCs}

Finally, we investigate the quantum geometrical properties of a $d$-wave SC within the context of mean field theory, which may be particularly relevant to the overdoped regime of the phase diagram\cite{Damascelli03,Keimer15}. The energy dispersion and the gap are parametrized by
\begin{eqnarray}
&&\varepsilon_{\bf k}=-2t(\cos k_{x}+\cos k_{y})+4t'\cos k_{x}\cos k_{y}-\mu=d_{3},
\nonumber \\
&&\Delta_{\bf k}=2\Delta_{0}(\cos k_{x}-\cos k_{y})=d_{1},
\end{eqnarray}
and $E_{\bf k}=\sqrt{\varepsilon_{\bf k}^{2}+\Delta_{\bf k}^{2}}$. For concreteness, we use (in units of eV) $t=0.15$, $t'=0.04$, $\mu=-0.13$, and $\Delta_{0}=0.1$ that are appropriate for optimally doped to slightly overdoped\cite{Nunner06} Bi$_{2}$Sr$_{2}$CaCu$_{2}$O$_{8+x}$. The quantum metric calculated from Eq.~(\ref{bare_quantum_metric}) takes the vielbein form
\begin{eqnarray}
&&g_{\mu\nu}=e_{\mu}e_{\nu},
\nonumber \\
&&e_{x}=\frac{\Delta_{0}}{E_{\bf k}^{2}}\sin k_{x}\left(4t\cos k_{y}-4t'\cos^{2}k_{y}+\mu\right),
\nonumber \\
&&e_{y}=\frac{\Delta_{0}}{E_{\bf k}^{2}}\sin k_{y}\left(-4t\cos k_{x}+4t'\cos^{2}k_{x}-\mu\right).
\end{eqnarray}
In Fig.~\ref{fig:2D_DSC_quantum_metric} (a), we present the unit vector field ${\bf w}_{\bf k}=(v_{\bf k},-u_{\bf k})$ of the quasihole state, which exhibits a vortex like feature near the nodal point ${\bf k}_{0}$ where the energy dispersion $E_{\bf k_{0}}=0$, meaning that the quasihole state as a unit vector in the Hilbert space rotates very dramatically near the nodal point. As a result, the quantum metric shown in Fig.~\ref{fig:2D_DSC_quantum_metric} (b) also displays a very singular behavior that has a pair of maxima around the nodal point.

To get a clear physical picture about the peculiar momentum profile of the metric, an analytical expression can be given for the pedagogical case when we manually turn off $t'=\mu=0$ such that the Fermi surface has a diamond shape and the nodal point is located at ${\bf k}_{0}=(\pi/2,\pi/2)$. In this case, the corresponding bare quantum matric $\overline{g}_{\mu\nu}$ expanded around the nodal point ${\bf k}={\bf k}_{0}+\delta{\bf k}$ takes the form
\begin{eqnarray}
&&\overline{g}_{xx}\approx\frac{\Delta_{0}^{2}\,t^{2}\delta k_{y}^{2}}{\left[(t^{2}+\Delta_{0}^{2})(\delta k_{x}^{2}+\delta k_{y}^{2})+2(t^{2}-\Delta_{0}^{2})\delta k_{x}\delta k_{y}\right]^{2}},
\nonumber \\
&&\overline{g}_{yy}\approx\frac{\Delta_{0}^{2}\,t^{2}\delta k_{x}^{2}}{\left[(t^{2}+\Delta_{0}^{2})(\delta k_{x}^{2}+\delta k_{y}^{2})+2(t^{2}-\Delta_{0}^{2})\delta k_{x}\delta k_{y}\right]^{2}},
\nonumber \\
&&\overline{g}_{xy}\approx\frac{-\Delta_{0}^{2}\,t^{2}\delta k_{x}\delta k_{y}}{\left[(t^{2}+\Delta_{0}^{2})(\delta k_{x}^{2}+\delta k_{y}^{2})+2(t^{2}-\Delta_{0}^{2})\delta k_{x}\delta k_{y}\right]^{2}},
\nonumber \\
\label{quantum_metric_DSC_expansion}
\end{eqnarray}
which matches fairly well with the numerical results. We see that approaching the nodal point $\left\{\delta k_{x},\delta k_{y}\right\}\rightarrow 0$, the bare quantum metric diverges. In addition, by changing to polar coordinates $\left(\delta k_{x},\delta k_{y}\right)=\left(k\cos\theta,k\sin\theta\right)$, the expansion in Eq.~(\ref{quantum_metric_DSC_expansion}) becomes
\begin{eqnarray}
&&\overline{g}_{xx}\approx\frac{1}{k^{2}}\times
\frac{\Delta_{0}^{2}t^{2}\sin^{2}\theta}{\left[(t^{2}+\Delta_{0}^{2})+(t^{2}-\Delta_{0}^{2})\sin 2\theta\right]^{2}},
\nonumber \\
&&\overline{g}_{yy}\approx\frac{1}{k^{2}}\times
\frac{\Delta_{0}^{2}t^{2}\cos^{2}\theta}{\left[(t^{2}+\Delta_{0}^{2})+(t^{2}-\Delta_{0}^{2})\sin 2\theta\right]^{2}},
\nonumber \\
&&\overline{g}_{xy}\approx\frac{1}{k^{2}}\times
\frac{-\Delta_{0}^{2}t^{2}\sin\theta\cos\theta}{\left[(t^{2}+\Delta_{0}^{2})+(t^{2}-\Delta_{0}^{2})\sin 2\theta\right]^{2}},
\end{eqnarray}
which after a polar integration $\int k\,dk\,d\theta\,\overline{g}_{\mu\nu}$ diverges logarithmically, indicating that the fidelity number in Eq.~(\ref{Gmunu_definition}) diverges for $d$-wave SCs, and therefore it may not be directly related to the electromagnetic responses we have discussed for $s$-wave SCs. This also implies that the average distance between Bloch states in the BZ of $d$-wave SCs diverges, owing to the very singular behavior near the nodal points. 

%As a result, the fidelity marker defined in Eq.~(\ref{fidelity_marker_definition}) should also diverge, so we omit to calculate the marker. 

%{\cblue (1) Need to check if d-wave BCS mean field theory gives the critical exponent $\beta=1/2$. If so then the fidelity number will vanish linearly with $T-T_{c}$. We can say that in reality, if the optical absorption measurement on cuprates reveals that the critical exponent of the absorption rate $2\beta\neq 1$, it would imply that the superconductivity is not entirely described by our weak coupling mean field theory, and maybe some other correlation effects need to be taken into account.   }

\subsection{Topological charge and metric-curvature correspondence in $d$-wave SCs}

%{\cblue (2) If cuprates and graphene are so similar in the sense of topological charge, then cuprates should also have some edge states if one orients it like the zigzag ribbon, no? Or just consider a ribbon of cuprates. ----- Yes, but Schnyder15 already points this out. }

\begin{figure}[ht]
\begin{center}
\includegraphics[clip=true,width=0.8\columnwidth]{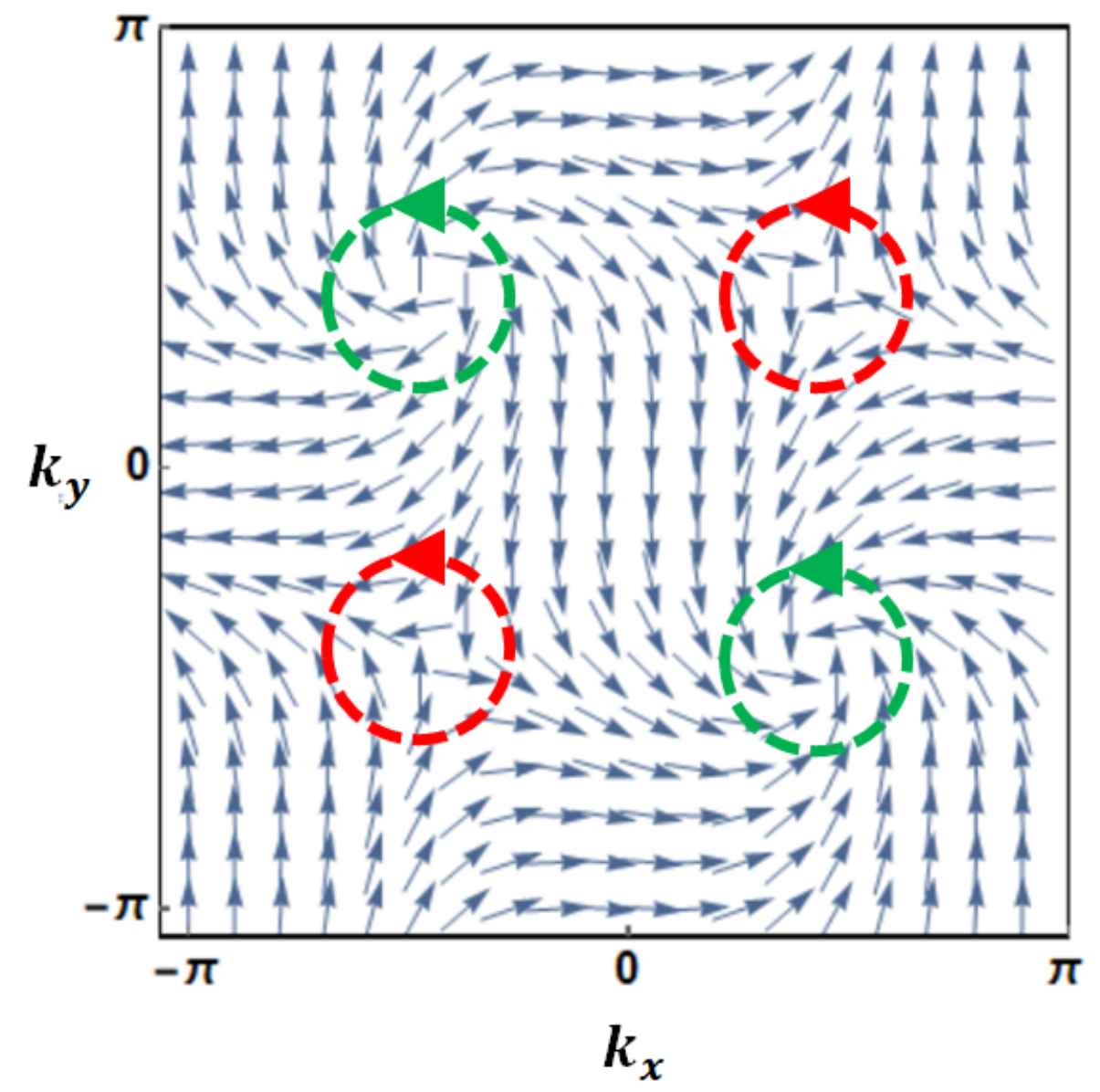}
\caption{The vector ${\bf n}=(n_{1},n_{3})$ defined in Eq.~(\ref{nvector_definition}) plotted as unit vectors in the momentum space of a $d$-wave SC. One sees that the vector field has a nonzero winding about each nodal point as indicated by the circles, and the winding is opposite between two pairs of nodal points (red circles and green circles have opposite winding numbers).  } 
\label{fig:DSC_n_field}
\end{center}
\end{figure}

The issue of topological charge in 2D $d$-wave SCs has been discussed within the context of nodal SCs, where it has been pointed out that the nodal points possess nonzero winding numbers\cite{Schnyder15,Le22}. In this section, we elaborate that the winding number can be visualized by the ${\bf n}$-field defined in Eq.~(\ref{nvector_definition}), and moreover it has a correspondence with the quantum metric. Our observation is that the non-Abelian Berry connection between quasihole and quasiparticle states of a singlet SC can generally be written as
\begin{eqnarray}
&&\langle m|\partial_{\mu}n\rangle=-{\rm sgn}(\Delta_{\bf k})\langle n|{\hat C}i\partial_{\mu}|n\rangle
\nonumber \\
&&=-{\rm sgn}(d_{1})\frac{d_{1}\partial_{\mu}d_{3}-d_{3}\partial_{\mu}d_{1}}{2d^{2}}
\nonumber \\
&&=-\frac{{\rm sgn}(\Delta_{\bf k})}{2}\left(n_{1}\partial_{\mu}n_{3}-n_{3}\partial_{\mu}n_{1}\right).
\end{eqnarray}
In the first line of this equation we have used the operator ${\hat C}=\sigma_{2}K$ that implements the particle-hole (PH) symmetry\cite{Le22} ${\hat C}H({\bf k}){\hat C}^{-1}=-H(-{\bf k})$, which indicates that the non-Abelian Berry connection can as well be implemented as a kind of charge-conjugated Berry connection that is dressed by the PH operator ${\hat C}$.
In a 2D $d$-wave SC, one can introduce a winding number that counts how many times the ${\bf n}$-vector winds along a circle enclosing a nodal point. The integration of the non-Abelian Berry connection weighted by the sign of the gap along such a circle is equivalently this winding number (not to be confused with the PH operator ${\hat C}$)
\begin{eqnarray}
&&{\cal C}=\oint\frac{d\phi}{2\pi}(n_{1}\partial_{\phi}n_{3}-n_{3}\partial_{\phi}n_{1})
\nonumber \\
&&=-2\oint\frac{d\phi}{2\pi}{\rm sgn}(\Delta_{\bf k})\langle m|\partial_{\phi}n\rangle
=2\oint\frac{d\phi}{2\pi}\langle n|{\hat C}i\partial_{\phi}|n\rangle
\nonumber \\
&&\equiv\oint\frac{d\phi}{2\pi}J_{\bf n},
\label{DSC_topo_charge}
\end{eqnarray}
where $\phi$ is the polar angle along the circle.
As shown in Fig.~\ref{fig:DSC_n_field}, by plotting the ${\bf n}$-vector in the momentum space, we see clearly that each nodal point corresponds to a nonzero winding number\cite{Le22} (or topological charge) ${\cal C}=\pm 1$. Furthermore, the integrand of this topological charge may be written as a determinant
\begin{eqnarray}
J_{\bf n}=\det\left({\bf n},\partial_{\phi}{\bf n}\right)
=\left|\begin{array}{cc}
n_{1} & \partial_{\phi}n_{1} \\
n_{3} & \partial_{\phi}n_{3} 
\end{array}\right|\equiv\det E_{\bf n}.
\end{eqnarray}
As a result, the square of the integrand is equal to the azimuthal quantum metric 
\begin{eqnarray}
&&|J_{\bf n}|^{2}=\det E_{\bf n}^{T}E_{\bf n}=\det\left(
\begin{array}{cc}
{\bf n}\cdot{\bf n} & {\bf n}\cdot\partial_{\phi}{\bf n} \\
\partial_{\phi}{\bf n}\cdot{\bf n} & \partial_{\varphi}{\bf n}\cdot \partial_{\varphi}{\bf n}
\end{array}
\right)
\nonumber \\
&&=\partial_{\varphi}{\bf n}\cdot \partial_{\varphi}{\bf n}=4\,g_{\phi\phi},
\end{eqnarray}
after using ${\bf n}\cdot{\bf n}=1$ and Eq.~(\ref{bare_quantum_metric}). This relation between the integrand of the topological charge and the quantum metric has been referred to as the metric-curvature correspondence, which is found to be true in any TIs and topological superconductors described by a Dirac model, as well as 2D Dirac semimetals like graphene. In this sense, graphene and $d$-wave SCs actually have very similar topological and quantum geometrical properties.

%provided one considers the topological charge constructed from the non-Abelian Berry connection in Eq.~(\ref{DSC_topo_charge}). 

%This correspondence also implies that if one is able to measure the momentum profile of $g_{\phi\phi}({\bf k})$, then the topological charge can be unambiguously identified, although it remains unclear to us at present how $g_{\phi\phi}({\bf k})$ can be measured in SCs.

\section{Conclusions}

In summary, we elaborate that the filled quasihole state $|n\rangle$ of singlet SCs possesses nontrivial quantum geometrical properties. The quantum metric defined from the overlap of quasihole states at momenta ${\bf k}$ and ${\bf k+\delta k}$ is nonzero, and can be simply understood as the twisting of the quasihole state as a unit vector in the Hilbert space that can be visualized from the Bogoliubov coefficients ${\bf w}_{\bf k}=(v_{\bf k},-u_{\bf k})$. In addition, the momentum integration of quantum metric yields a nonzero fidelity number, which is a measure of average distance between neighboring quasihole states in the BZ, and equivalently the spread of quasihole Wannier functions. For $s$-wave SCs, the fidelity number is essentially the coherence length measured in terms of the lattice constant and then multiplied by the correct unit. In other words, the coherence length is actually a measure of the quantum geometry in s-wave SCs. We further show that the quantum metric and fidelity number enter various electromagnetic responses such as infrared absorption, paramagnetic current, and dielectric function, indicating that these responses are directly related to the quantum geometry. The fidelity number can be further defined on lattice sites as a fidelity marker, and we find that nonmagnetic impurities locally suppress the marker, signifying the influence of disorder on the quantum geometrical properties of the $s$-wave SC. In contrast, for $d$-wave SCs, we find that the quantum metric exhibits a very singular profile near the nodal points, rendering a divergent fidelity number. Besides, the non-Abelian Berry connection that integrates to a topological charge of the nodal points is actually equivalent to the azimuthal quantum metric, satisfying a metric-curvature correspondence. Our theory thus clarifies the quantum geometrical properties of singlet SCs, the possibility of measuring them experimentally, as well as how disorder may influence these properties. Many related issues, such as whether the same aspects also applies to triplet SCs of various pairing symmetries, await to be explored.

%i.e., the average distance between neighboring quasihole state diverges. As a result, 

\begin{acknowledgments}

W. C. acknowledges the financial support from the productivity in research fellowship from CNPq, and D. P. is supported by the Mestrado Nota 10 fellowship from FAPERJ.

%and D. P. acknowledges the support from the FAPERJ Nota 10 fellowship.

\end{acknowledgments}

\bibliography{Literatur}

\end{document}